\documentclass[12pt]{article}

\usepackage{amsfonts}
\usepackage{latexsym}

\newcommand{\T}{\vartheta}
\newcommand{\PHI}{\vartheta^{\dagger}}

\newcommand{\PP}{{\mathcal P}}
\newcommand{\QQ}{{\mathcal Q}}
\newcommand{\dX}{{\mathcal X}}
\newcommand{\dY}{{\mathcal Y}}
\newcommand{\dA}{{\mathcal A}}
\newcommand{\dB}{{\mathcal B}}
\newcommand{\dC}{{\mathcal C}}
\newcommand{\HH}{{\mathcal M}}
\newcommand{\dH}{{\bf H}}
\newcommand{\dK}{{\bf K}}

\newcommand{\dHH}{{\mathcal H}}
\newcommand{\dKK}{{\mathcal K}}
\newcommand{\Dk}{\mathfrak{k}}
\newcommand{\Dh}{\mathfrak{h}}
\newcommand{\dkk}{\kappa}
\newcommand{\dhh}{n}

\newcommand{\Di}{\Delta_{(1)}}
\newcommand{\Dj}{\Delta_{(2)}}
\newcommand{\Dij}{\Delta_{(1)}\Delta_{(2)}}

\newcommand{\N}{{\mathcal N}}

\newcommand{\psiu}{\psi^{\uparrow}}
\newcommand{\psid}{\psi^{\downarrow}}

\def\mref#1{(\ref{#1})}
\def\eqref#1{(\ref{#1})}

\begin{document}
\title{A Laplace ladder of discrete Laplace equations}
\author{Maciej Nieszporski\\
\\
Instytut Fizyki Teoretycznej,
Uniwersytet w Bia{\l}ymstoku,\\
ul. Lipowa 41, 15-424 Bia{\l}ystok, Poland\\
 e-mails: maciejun@fuw.edu.pl, maciejun@alpha.uwb.edu.pl \\
tel: 48-85-7457239, fax: 48-85-7457238
}
\date{}
\maketitle

\begin{abstract}
The notion of a Laplace ladder for a discrete analogue of the 
 Laplace equation is presented.
The adjoint of the discrete Moutard equation and a discrete  counterpart of
the nonlinear  form of Goursat equation are introduced.
 \end{abstract} 

\section{Notation}
Value of functions of continuous variables we denote by $f(u,v)$
i.e. 
$f:{\mathbb R}^2 \ni (u,v) \mapsto  f(u,v) \in {\mathbb R}$ while
functions of discrete variables  by 
$f(m_1,m_2)$
i.e. 
$f:{\mathbb Z}^2 \ni (m_1,m_2) \mapsto f(m_1,m_2) \in {\mathbb R} $.
Partial derivatives are
 denoted by comma e.g. 
$f,_{uv}(u,v):=\frac{\partial^2 f}{\partial u \partial v}(u,v)$.
Shift operators are denoted by
subscripts in brackets e.g. $f(m_1,m_2)_{(1)}:=f(m_1+1,m_2)$,
$f(m_1,m_2)_{(2)}:=f(m_1,m_2+1)$, 
$f(m_1,m_2)_{(12)}:=f(m_1+1,m_2+1)$, $f(m_1,m_2)_{(-1)}:=f(m_1-1,m_2)$ etc.
We omit arguments when operators indicate what kind of functions
(of discrete 
or  continuous variables) we deal with.
Difference operators are denoted by $\Delta_i f:=f_{(i)}-f$.
The diamond operator we define as follows $\Diamond f:=\frac{f_{(12)}f}{f_{(1)}f_{(2)}}$.  

\section{Introduction}
The Laplace equation 
\begin{equation}
\label{LaplaceE1}
\begin{array}{l}
L \psi(u,v)=0
\end{array}
\end{equation}
where $L$ is a differential operator
 \begin{equation}
\begin{array}{l}
L=\frac{\partial^2}{\partial u \partial v}-
A (u,v) \frac{\partial}{\partial u} 
-B (u,v) \frac{\partial}{\partial v}-C (u,v),
\end{array} 
\end{equation}
is often regarded as a master equation
of theory of $S$-integrable systems.
One can justify this point of view 
indicating on either   the existence of 
large class of Darboux-type transformations introduced by 
Jonas \cite{JonasF} and Eisenhart 
\cite{Eisenhart} 
(called  fundamental transformation or Darboux-type transformation
for multicomponent KP hierarchies) 
 that acts on  solution spaces of set of the Laplace equations
and give rise Darboux-B\"acklund transformations  for the 
Darboux equations or 
modern approach  of $\bar{\partial}$-dressing  
method \cite{Zakharov} that was applied 
to Darboux equations by Zakharov and Manakov. 

Both $\bar{\partial}$-dressing  method and, what more important for us,
discrete version of fundamental
transformation acting on   solution spaces  of discrete counterparts 
of the Laplace equations
\begin{equation}
\label{DLaplaceE}
\begin{array}{l}
{\mathcal L} \psi(m_1,m_2)=0 
\end{array}
\end{equation}
\begin{equation}
\label{DLEO}
\begin{array}{l}
 {\mathcal L}=\Dij   - \dA(m_1,m_2) \Di 
-\dB(m_1,m_2) \Dj -
 \dC(m_1,m_2) , 
\end{array}
\end{equation}
was  applied to discrete analogues of  Darboux equations
\cite{BK,Manas,DSM}. Moreover a theory of reductions of the discrete fundamental 
transformation is being developed 
\cite{Nimmo,CDS,KS,DMS,DSSYM,DoliwaQ}.
 
On the other hand Athorne \cite{Athorne} indicated that a symmetry of a Laplace
ladder of Laplace equations impose  constraints on 
the operators $L$ of equations of the ladder. In particular the cases 
when Laplace ladder contains a Moutard equation
\begin{equation}
\begin{array}{l}
\psi,_{uv}=f(u,v) \psi
\end{array}
\end{equation}
and Goursat equation
\begin{equation}
\begin{array}{l}
\psi,_{uv}=\frac{1}{2} (\log \lambda (u,v)),_v  \psi,_u+\lambda (u,v) \, \psi
\end{array}
\end{equation}
which is related to \cite{Goursat,Ganhza}
\begin{equation}
\label{NG}
\begin{array}{l}
\vartheta_{xy}= 2 \sqrt{\lambda(x,y) \vartheta_{x} \vartheta_{y}}
\end{array}
\end{equation}
were discussed.
 There exist Darboux type transformations
that leave the form  of Moutard equation \cite{Moutard} and Goursat equation
\cite{Goursat,Ganhza} unchanged. 

We start from the basic facts from the theory of discrete Laplace equation
(the invariant description and $T$-equivalence are introduced
in section \ref{LAPLACE}). Next we introduce two operations
acting on invariants of discrete Laplace equation:
 Laplace transformations
(following the papers \cite{DoliwaL,Adler,Novikov} we recall
Laplace sequence of discrete Laplace equations ) 
in section \ref{LAPLACET} and an adjoint operation
 in section \ref{FUNDAMENTAL}.
On combining the notions of Laplace sequence and adjoint operation
we introduce in section \ref{LADDER} Laplace 
ladder of discrete Laplace equations.  
The Laplace ladder turns out to be a proper object
to  discuss an old and a new discrete Moutard type equation that respectively 
appear
first in the papers  \cite{Nimmo} and \cite{NDS,DNS} (see section \ref{MOUTARD})
and the discrete Goursat equation \cite{DSSYM} as well (see section \ref{GOURSAT}).

\section{Discrete Laplace equation. $T$-equivalence.}
\label{LAPLACE}
A differential operator \mref{LaplaceE1}
does not change its form under the transformation
\begin{equation}
\label{cech}
\begin{array}{l}
 L \mapsto \tilde{L} =\frac{1}{g} \circ L \circ g
\end{array} 
\end{equation}
where  $g$ denotes an  operator of multiplying by a scalar function
\hbox{$g=g(u,v)\ne 0$}.
The above transformation is usually called
{\em gauge transformation}. 
We say that  two {\em Laplace equations} 
 $L \psi(u,v)=0$
and $\tilde{L} \psi(u,v)=0$ are {\em equivalent},
{\em iff} there exists a function $g$ such that operators $\tilde{L}$ and $L$ 
are related by $\mref{cech}$
(Relation: \{two Laplace equation $L \psi(u,v)=0$
and $\tilde{L} \psi(u,v)=0$ are in the relation {\em iff} 
there exist a gauge 
transformation $\mref{cech}$\} is equivalence relation). 
The following functions ({\em Laplace invariants})
are invariant with respect to  the gauge transformation
\begin{equation}
\label{hk}
\begin{array}{l}
h=AB-A ,_u +C, \qquad
k=AB-B ,_v +C. 
\end{array}
\end{equation}
If the invariants of the operators $L$ and $\tilde{L}$ are equal
($k=\tilde{k}$ and $h=\tilde{h}$) 
then there exist a gauge transformation that transforms 
$\tilde{L}$ into  ${L}$.
Ordered pair of invariants $(h,k)$ labels the equivalence classes
of equations  \mref{LaplaceE1}.
We refer to Athorne paper
\cite{Athorne} for further information concerning the continuous case.

 The  discrete Laplace equation \mref{DLaplaceE} 
 can be written in the form
\begin{equation}
\label{DLaplaceE1}
\begin{array}{c}
\psi _{(12)}-\alpha \psi _{(1)}- \beta \psi _{(2)}- \gamma \psi=0 
\end{array}
\end{equation}
where 
\begin{equation}
\begin{array}{c}
\alpha:=\dA+1 \qquad \beta:=\dB+1 \qquad \gamma:=\dC-\dA-\dB-1
\end{array}
\end{equation}
By analogy to the continuous case we introduce  invariants
\begin{equation}
\begin{array}{l}
\dkk:=\frac{ \beta \alpha_{(2)} }{ \gamma _{(2)} },   \qquad
\dhh:=\frac{ \alpha \beta _{(1)}}{ \gamma _{(1)} }. 
\end{array}
\end{equation}
with respect to a discrete gauge transformations ${\mathcal L} \mapsto \frac{1}{g _{(12)}} {\mathcal L} g $,
which is the discrete counterpart of the gauge transformation
 \mref{cech}
and where $g$ denotes 
operator of multiplication by function
$g=g(m_1,m_2)\ne 0$.
We shall call the invariants {\em basic invariants}.
They are basic in the sense that if two operators ${\mathcal L}$ and $\tilde{{\mathcal L}}$
have equal  basic invariants ($(\dhh,\dkk)=(\tilde{\dhh},\tilde{\dkk})$)
then they can be  related by a gauge.

For our convenience (section \ref{MOUTARD}) 
we introduce also derivative invariants which do not poses
the basic property.
{\em
Secondary
invariants of the first kind} are defined by
\begin{equation}
\begin{array}{l}
\dKK:=\dkk \dkk_{(1)} \qquad
\dHH:=\dhh \dhh_{(2)}
\end{array}
\end{equation}
{\em 
Secondary 
invariants of the first kind} are defined by
\begin{equation}
\begin{array}{l}
\dK:=\frac{\dhh _{(2)}}{\dkk} \qquad
\dH:=\frac{\dkk _{(1)}}{\dhh}
\end{array}
\end{equation}
We denote  equivalence
classes of the discrete Laplace equations by $(\dhh,\dkk)$.

If the basic invariants of two discrete Laplace operators 
${\mathcal L}$ and  $\tilde{{\mathcal L}}$ are related as follows
\[\dkk=T \tilde{\dkk}, \qquad \dhh=T \tilde{\dhh},\] 
where $T$ denotes a shift
operator (i.e. $T f(m_1,m_2)=f(m_1+k,m_2+l)$ 
where $k\in{\mathbb N}$ and  $l\in{\mathbb N}$)
then we shall say that equations are {\em $T$-equivalent}  
and we denote the enlarged equivalence classes by $T(\dhh,\dkk)$.
 $T$-equivalence plays important role in our considerations.

\section{Laplace transformations and Laplace sequence}
\label{LAPLACET}
In order to solve   a discrete Laplace equation \mref{DLaplaceE}
we can try to factorize the discrete
Laplace operator \mref{DLEO}. There are only two ways we can do it
\begin{eqnarray}
\label{dfact1}
 (\Di-\dB) (\Dj-\dA_{(-1)})
\psi - {\Dh}\psi=0 ,\\
\label{dfact2}
 (\Dj-\dA) (\Di-\dB_{(-2)})
\psi -{\Dk} \psi=0, 
\end{eqnarray}
where 
\begin{equation}
\begin{array}{l}
\Dh=\dC-\dA+\dA_{(-1)}(\dB+1)=\gamma+\alpha _{(-1)} \beta , \\
\Dk=\dC-\dB+\dB_{(-2)}(\dA+1)=\gamma+\beta _{(-2)} \alpha .
\end{array}
\end{equation}
If one of the functions $\Dh$ or $\Dk$ is equal to zero we succeed in factorization (and we can
solve equation by quadratures).
If not we can introduce functions
\begin{eqnarray}
\label{dfact3}
\psiu:=(\Dj  -\dA_{(-1)}) \psi\\
\label{dfact4}
\psid:=(\Di -\dB_{(-2)}) \psi
\end{eqnarray}
equations  \mref{dfact1} and \mref{dfact2} can be written in the form
\begin{eqnarray}
\label{dfact5}
\psi=\frac{1}{\Dh}(\Di \psiu-\dB \psiu)\\
\label{dfact6}
\psi=\frac{1}{\Dk}(\Dj \psid-\dA \psid)
\end{eqnarray}
If we eliminate from equations (\ref{dfact3}--\ref{dfact4}) the 
function $\psi$ 
using  (\ref{dfact5}--\ref{dfact6}) then we obtain that functions
$\psiu$ and $\psid$  satisfy equations
\begin{eqnarray}
\label{psiu}
\psiu _{(12)} = \frac{\Dh_{(2)}}{\Dh} \alpha_{(-1)} \psiu _{(1)} +
\beta _{(2)} \psiu _{(2)}+\frac{\Dh_{(2)}}{\Dh} \gamma \psiu\\ 
\psid _{(12)} =  \alpha_{(1)} \psid _{(1)} +
\frac{\Dk_{(1)}}{\Dk} \beta _{(-2)} \psid _{(2)}+
\frac{\Dk_{(1)}}{\Dk} \gamma \psid
\end{eqnarray}
and invariants of these equations  are connected with
invariants of the equation \mref{DLaplaceE1}
via
\begin{equation}
\label{dTLp}
\begin{array}{ll}
\dhh^{\uparrow}= 
\frac{\dhh_{(-1)} \dhh_{(2)}}{\dkk} \Diamond \frac{1}{(1+\dhh)_{(-1)}} &\qquad
\dkk^{\uparrow}_{(1)}=\dhh_{(2)}  \\
\dhh^{\downarrow}_{(2)}=\dkk_{(1)} &\qquad
\dkk^{\downarrow}=
\frac{\dkk_{(-2)} \dkk_{(1)}}{\dhh} \Diamond \frac{1}{(1+\dkk)_{(-2)}} 
\end{array}
\end{equation}
\begin{equation}
\label{dTLn}
\begin{array}{ll}
\dH^{\uparrow}= \left( \dH \Diamond (1+\dhh) \right)_{(-1)} &\qquad
\dK^{\uparrow}_{(1)}=\left( \dK_{(1)} \Diamond \frac{1}{(1+\dhh)} \right)_{(2)} \\
\dH^{\downarrow}_{(2)}=
\left(  \dH_{(2)} \Diamond \frac{1}{(1+\dkk)} \right) _{(1)} &\qquad
\dK^{\downarrow}=\left( \dK \Diamond (1+\dkk) \right) _{(-2)} 
\end{array}
\end{equation}
One can easily show, that
\[(\dhh^{\uparrow\downarrow},\dkk^{\uparrow\downarrow})=
(\dhh^{\downarrow\uparrow},\dkk^{\downarrow\uparrow})=(\dhh,\dkk)\]
so {\em Laplace transformations} for the discrete Laplace equations i.e. maps
$(\dhh,\dkk) \mapsto (\dhh^{\uparrow},\dkk^{\uparrow})$ and 
$(\dhh,\dkk) \mapsto (\dhh^{\downarrow},\dkk^{\downarrow})$ are mutually
inverse. By analogy to continuous case we call the sequence of the equations
\begin{equation}
\label{dcL}
...,(\dhh^{\downarrow\downarrow},\dkk^{\downarrow\downarrow}),
(\dhh^{\downarrow},\dkk^{\downarrow}),(\dhh,\dkk),(\dhh^{\uparrow},\dkk^{\uparrow}),
(\dhh^{\uparrow\uparrow},\dkk^{\uparrow\uparrow}),...
\end{equation}
  {\em  Laplace sequence of discrete Laplace equations}. 
Three comments are in order.
Firstly, Laplace transformation are well defined on the $T$-equivalence classes.
Secondly equation
\[\dhh^{\uparrow}_{(1)} \dhh^{\downarrow}_{(2)}=
\dhh \dhh_{(12)} \Diamond \frac{1}{1+\dhh}\]
where the Laplace transformation $\uparrow$  ($\downarrow$) are treated
as increment (decrement) in third discrete variable is nothing but 
a form of Hirota equation \cite{DoliwaH}.
Finally if we introduce $\dX:=\frac{\psi^{\uparrow}}{b}$ 
and $\dY:=\frac{\psi}{a}$ where $a$ and $b$ are a solutions of $b_{(1)}=\beta b$
$a_{(2)}=\alpha_{(-1)} a$ then equations  \mref{dfact3}
and \mref{dfact5} take form
\begin{equation}
\label{DrkpsG}
\begin{array}{l}
\Dj \dY=\QQ \dX\\
\Di \dX=\PP \dY
\end{array}
\end{equation}
where 
\begin{equation}
\label{PQ}
\PP =\frac{a \Dh}{b_{(1)}} \qquad \QQ=\frac{b}{a_{(2)}}
\end{equation}
and equations \mref{psiu} and \mref{DLaplaceE1}  in the new gauge are
\begin{eqnarray}
\label{dkpsG}
\label{Y}
\dY _{(12)}=
\dY _{(1)}+\frac{\QQ _{(1)}}{\QQ}\dY _{(2)}-\HH\frac{\QQ _{(1)}}{\QQ}\dY\\
\label{X}
\dX _{(12)}=
\frac{\PP _{(2)}}{\PP} \dX _{(1)}+\dX _{(2)}-\HH\frac{\PP _{(2)}}{\PP}\dX
\end{eqnarray}
where \[\HH =1-\PP\QQ.\]

\section{Fundamental transformation and adjoint}
\label{FUNDAMENTAL}
Fundamental transformation that maps from solution space of
equation  \mref{LaplaceE1}
has two  functional parameters \cite{Eisenhart}. The first one is 
a solution of equation \mref{LaplaceE1} and the second one is a
 solution of adjoint of  equation \mref{LaplaceE1}
\begin{equation}
\label{sprz}
 \psi^{\dagger},_{uv}  
+\left( A \psi^{\dagger} \right),_u + 
\left( B \psi^{\dagger} \right),_v - C \psi^{\dagger} = 0. 
\end{equation}
Our aim is to introduce the discrete  adjoint of
equation \mref{DLaplaceE1}.
To do it we shall  suitable  rewrite fundamental transformation
\cite{DSM} for
the single discrete Laplace equation  \mref{LaplaceE1}.

Discrete fundamental transformation maps from the solution space of the equation
(we choose  for convenience the so called affine gauge 
i.e. $\mathcal C=0$)
\begin{equation}
\label{Dconj}
\Di \Dj x = \frac{\Dj a}{a} \Di x +\frac{\Di b}{b} \Dj x 
\end{equation} 
to  the  solution space of the equation
\begin{equation}
\label{Dconjp}
\Di \Dj x^1 = \frac{\Dj a^1}{a^1} \Di x^1 +\frac{\Di b^1}{b^1} \Dj x^1 
\end{equation} 
and is given by
\begin{equation}
\label{DtranF}
\begin{array}{l}
\frac{1}{\Di (\frac{ \T '}{\T})} \Di (\frac{x^1 \T '}{\T})=
\frac{1}{\Di (\frac{1}{\T})} \Di(\frac{x}{\T})\\
\frac{1}{\Dj (\frac{ \T '}{\T})} \Dj (\frac{x^1 \T '}{\T})=
\frac{1}{\Dj (\frac{1}{\T})} \Dj (\frac{x}{\T}).
\end{array}
\end{equation}
The functional parameters
of the 
transformation  $\T$ i $\T '$ are not arbitrary, namely
the function  $\T$ is a solution of
equation \mref{Dconj} i.e.
\begin{equation}
\label{DtranF1}
\Di \Dj \T = \frac{\Dj   a}{a} \Di \T  +\frac{\Di  b}{b} \Dj \T 
\end{equation}
while  function $\T '$
we construct from  $\T$   and a solution
 $\PHI$ of the equation
\begin{equation}
\label{DtranF2}
\Di \Dj \PHI  +\Di \left( \frac{\frac{\Dj a}{a}\PHI}{1+\frac{\Dj a}{a}+\frac{\Di b}{b}} \right)  
+ \Dj  \left( \frac{\frac{\Di b}{b}\PHI}{1+\frac{\Dj a}{a}+\frac{\Di b}{b}} \right)   = 0 
\end{equation}
The way to obtain  $\T '$ looks as follow: first we find auxiliary functions
  $\lambda$ i $\chi$
\begin{equation}
\label{DtranF3}
\begin{array}{lll}
\Di \lambda = \frac{\frac{\Di b}{b}\PHI}{1+\frac{\Dj a}{a}+\frac{\Di b}{b}} &&
\Dj \lambda =-\Dj \PHI-\frac{\frac{\Dj a}{a}\PHI}{1+\frac{\Dj a}{a}+\frac{\Di b}{b}}  \\
&\chi = \lambda + \PHI & 
\end{array}
\end{equation}
and next  function  $\T '$ we are searching for
\begin{equation}
\label{DtranF4}
\begin{array}{l}
\Di \T' = \chi \Di \T,\\
\Dj \T' = \lambda \Dj \T.
\end{array}
\end{equation} 
Conditions (\ref{DtranF1}-\ref{DtranF4}) 
assure us, that $x^1$ given by \mref{DtranF} satisfies
 \mref{Dconjp}. New functions $a^1$ and $b^1$ are related to old ones via 
\begin{equation}
\begin{array}{lll}
\frac{\Dj a^1}{a^1} =  
\frac{ \Dj ( a \chi \frac{\T}{\T'} - a)}{ ( a \chi \frac{\T}{\T'} - a)} & , &
\frac{\Di b^1}{b^1} =  
\frac{\Di ( b \lambda \frac{\T}{\T'} - b)}{( b \lambda \frac{\T}{\T'} - b)}
\end{array}
\end{equation}
We say that equation \mref{DtranF2} is adjoint of equation \mref{DtranF1}.  
Invariants $\dhh^{\dagger}$, $\dkk^{\dagger}$, $\dH^{\dagger}$ and
$\dK^{\dagger}$  of equation 
\mref{DtranF2} are conected with   invariant
$\dhh$, $\dkk$, $\dH$ and  $\dK$ of equation \mref{DtranF1}
in the following way
\begin{equation}
\label{indag}
\begin{array}{l}
\dhh^{\dagger}=\dkk_{(1)} \qquad \dkk^{\dagger}=\dhh_{(2)},
\end{array}
\end{equation}
\begin{equation}
\label{dsprzn}
\begin{array}{l}
\dH^{\dagger}=\dK_{(1)} \qquad \dK^{\dagger}=\dH_{(2)}.
\end{array}
\end{equation}
We say that equation ${\mathcal L}^{\dagger} \psi^{\dagger} =0$ is adjoint of equation
${\mathcal L} \psi=0$
{\em iff} 
\begin{equation}
\begin{array}{l}
T(\dhh^{\dagger},\dkk^{\dagger})=T(\dkk_{(1)},\dhh_{(2)}).
\end{array}
\end{equation}
 We observe that
$T(\dhh,\dkk)^{\dagger\dagger}=T(\dhh,\dkk)$.
\section{Laplace ladder}
\label{LADDER}
On superposing Laplace and adjoint transformations we get
\begin{equation}
\begin{array}{l}
(\dhh^{\uparrow \dagger},\dkk^{\uparrow \dagger})=
\left( \frac{\dhh{(-1)} \dhh{(2)}}{\dkk} \Diamond \frac{1}{1+\dhh_{(-1)}}, \dhh \right)_{(2)}\\
(\dhh^{ \dagger\downarrow},\dkk^{\dagger\downarrow})=
\left( \frac{\dhh{(-1)} \dhh{(2)}}{\dkk} \Diamond \frac{1}{1+\dhh_{(-1)}}, \dhh \right)_{(1)}\\
(\dhh^{\downarrow \dagger},\dkk^{\downarrow \dagger})=
\left( \frac{\dkk{(-2)} \dkk{(1)}}{\dhh} \Diamond \frac{1}{1+\dkk_{(-2)}}, \dkk \right)_{(1)}\\
(\dhh^{ \dagger\uparrow},\dkk^{\dagger\uparrow})=
\left( \frac{\dkk{(-2)} \dkk{(1)}}{\dhh} \Diamond \frac{1}{1+\dkk_{(-2)}}, \dkk \right)_{(2)}
\end{array}
\end{equation}
It means that we have
\begin{equation}
\begin{array}{l}
T(\dhh^{\uparrow \dagger},\dkk^{\uparrow \dagger})=
T(\dhh^{\dagger \downarrow},\dkk^{\dagger \downarrow}), \qquad
T(\dhh^{ \downarrow\dagger},\dkk^{ \downarrow\dagger})=
T(\dhh^{ \dagger\uparrow},\dkk^{\dagger\uparrow}).
\end{array}
\end{equation}
We obtain a commuting diagram {\em Laplace ladder of discrete Laplace equations}
\small
\begin{displaymath}
\begin{array}{ccccccccc}
... 
&T(\dhh^{\downarrow},\dkk^{\downarrow}) & \longrightarrow &T(\dhh,\dkk)&
 \longrightarrow
&T(\dhh^{\uparrow},\dkk^{\uparrow}) &\longrightarrow 
&T(\dhh^{\uparrow\uparrow},\dkk^{\uparrow\uparrow})& ...\\
 & |  && | & & |  &&|&\\
...  &
T(\dhh^{\dagger\uparrow},\dkk^{\dagger\uparrow}) &\longleftarrow 
&T(\dhh^{\dagger},\dkk^{\dagger}) &
\longleftarrow &
 T(\dhh^{\dagger\downarrow},\dkk^{\dagger\downarrow}) & \longleftarrow &
T(\dhh^{\dagger\downarrow\downarrow},\dkk^{\dagger\downarrow\downarrow}) & ...
\end{array}
\end{displaymath}
\normalsize
where the long arrows denote $\uparrow$ Laplace transformations while vertical lines
(rungs of the ladder)
the adjoint operation ${\dagger}$.

\section{Moutard subcase}
\label{MOUTARD}
A Laplace equation is called {\em  discrete  Moutard equation}
{\em iff} its invariants are related by
\begin{equation}
\dH=\dK, 
\end{equation}
or equivalently $\dHH=\dKK$ or $\dhh_{(2)} \dhh=\dkk_{(1)} \dkk$.
Every discrete Moutard equation can be transformed to equation 
\begin{equation}
\label{dM}
N_{(12)} + N = F (N_{(1)} + N_{(2)}), 
\end{equation}
by gauge transformation.
The equation \mref{dM} together with the Darboux type transformation
has been introduced 
by Nimmo and Schief  \cite{Nimmo}.

From formulas \mref{dsprzn} we get that an invariant characterization of
the adjoint of discrete Moutard equation  is
\begin{equation}
\label{MoutardS}
\begin{array}{l}
\dH^{\dagger}_{(2)}=\dK^{\dagger}_{(1)}.
\end{array}
\end{equation}
So opposite to the continuous case the  discrete Moutard equation  
{\bf is not} self-adjoint.
A Laplace equation is called {\em adjoint discrete  Moutard equation}
{\em iff} its invariants are related by \mref{MoutardS}.
Every adjoint discrete Moutard equation can be transformed by a gauge transformation 
to equation 
\begin{equation}
\label{adM}
\N_{(12)} + \N = F_{(1)} \N_{(1)} + F_{(2)} \N_{(2)}. 
\end{equation}

If function  $N$ satisfies a discrete Moutard equation \mref{dM}
then function $\N$ given by
\begin{equation}
\label{cN}
\N:=N_{(1)}+N_{(2)} 
\end{equation}
satisfies the  adjoint discrete Moutard equation \mref{adM}.
 Due to relationship \mref{cN} 
equation \mref{adM} "inherit"  integrable features from equation
\mref{dM}.
In particular from the discrete Moutard transformation  \cite{Nimmo} we have
\begin{equation}
\begin{array}{l}
(\N '\frac{\Theta_{(1)}}{\Theta} +\N )_{(1)}=
\frac{(\Theta_{(1)}+\Theta_{(2)})_{(1)}}{\Theta_{(1)}+\Theta_{(2)}}
(\N '\frac{\Theta_{(2)}}{\Theta_{(12)}} +\N )
\\
(\N '\frac{\Theta_{(2)}}{\Theta} -\N )_{(2)}=
\frac{(\Theta_{(1)}+\Theta_{(2)})_{(2)}}{\Theta_{(1)}+\Theta_{(2)}}
(\N '\frac{\Theta_{(1)}}{\Theta_{(12)}} -\N )
\end{array}
\end{equation}
where  $\Theta$  is solution of discrete Moutard equation \mref{dM}
\[\Theta_{(12)}+\Theta=F(\Theta_{(1)}+\Theta_{(2)}).\]
Further features of the equation \mref{adM} will be discussed in
\cite{DoliwaK}.

\section{Goursat (symmetric) subcase}
\label{GOURSAT}
On demanding
 \begin{equation}
\label{dGou}
\begin{array}{l}
\dhh^{\uparrow}=\dkk 
\end{array}
\end{equation}
or equivalently 
\begin{equation}
\begin{array}{l}
\label{dGou1}
\dH^{\uparrow}=\dK
\end{array}
\end{equation}
the equations \mref{DLaplaceE1} and \mref{psiu},
which are equivalent respectively to equations \mref{Y} and 
 \mref{X}, 
become the {\em discrete  Goursat equations}.
Invariant characterization of the discrete  Goursat equations is
\begin{equation}
\begin{array}{l}
(\dhh^{\uparrow}_{(2)})^2=\dkk^{\uparrow}\dkk^{\uparrow}_{(12)} 
\Diamond \frac{1}{1+\dkk^{\uparrow}}\\
(\dkk_{(1)})^2=\dhh\dhh_{(12)} 
\Diamond \frac{1}{1+\dhh}
\end{array}
\end{equation}
In terms of functions given in \mref{PQ} we have \cite{DSSYM}
\begin{equation}
\label{dwsym}
\begin{array}{l}
\Diamond \frac{{\PP}}{{\QQ}}=\frac{\HH _{(2)}}{\HH _{(1)}}.
\end{array}
\end{equation}
On introducing function $\tau$
\begin{equation}
\begin{array}{l}
\frac{\tau_{(12)} \tau}{\tau_{(1)} \tau_{(2)}}=\HH=1-\PP \QQ
\end{array}
\end{equation}
and due to the fact that the equations \mref{DrkpsG} has the symmetry
\begin{displaymath}
\begin{array}{l}
(\dX,\dY,\PP,\QQ) \rightarrow 
\left(\dX V(m_2),\dY U(m_1),\PP\frac{U(m_1)}{V(m_2)},\QQ\frac{V(m_2)}{U(m_1)}\right)
\end{array}
\end{displaymath}
without lost of generality we can take
\begin{equation}
\begin{array}{l}
\PP \tau_{(1)}=\QQ \tau_{(2)}.
\end{array}
\end{equation}
On defining
\begin{equation}
\begin{array}{l}
x:=\dX \sqrt{\frac{\tau}{\tau_{(2)}}}, \qquad 
y:=\dY \sqrt{\frac{\tau}{\tau_{(1)}}},
\end{array}
\end{equation}
we get \cite{SchiefR} 
\begin{equation}
\label{drkpsG}
\begin{array}{l}
x_{(1)}= \frac{1}{\sqrt{\HH}}
\left(x+\PP \sqrt{\frac{\tau_{(1)}}{\tau_{(2)}}} \, y\right) ,  \qquad 
y_{(2)}= \frac{1}{\sqrt{\HH}}
\left(y+\QQ \sqrt{\frac{\tau_{(2)}}{\tau_{(1)}}} \, x\right)
.
\end{array}
\end{equation}
If functions $x$ and $y$ satisfy the above system then 
there exists a function $\theta$ such that  
\begin{equation}
\begin{array}{l}
\label{xyt}
 x^2= \Delta_{(2)} \theta  \qquad y^2= \Delta_{(1)} \theta
 \end{array}
\end{equation}
We also have
\begin{displaymath}
\begin{array}{l}
1-\HH=\PP^2 \frac{\tau_{(1)}}{\tau_{(2)}}=\QQ^2 \frac{\tau_{(2)}}{\tau_{(1)}}
\end{array}
\end{displaymath}
 and therefore squared equations \mref{drkpsG} (discrete analogue of \mref{NG})
take form
\begin{equation} 
\label{dRG}
\Dij \theta = \frac{1-\HH}{\HH} (\Di \theta+\Dj \theta)
\pm 2 \sqrt{\frac{1-\HH}{\HH^2}} \sqrt{\Di \theta}\sqrt{\Dj \theta} 
\end{equation}
where $\pm$ is result of square rooting of \mref{xyt}.
The equation \mref{dRG}  we shall call
{\em a nonlinear version   of discrete Goursat equation}.

In the case of Goursat reduction just as in the continuous case the ladder twist and
fold to the diagram
\begin{displaymath}
\begin{array}{ccccccccccc}
T(\dhh^{\dagger},\dkk^{\dagger}) & \longrightarrow & \cdot & \longrightarrow &
\cdot  & \longrightarrow
& \cdot &\longrightarrow &\cdot & ...\\
\uparrow | & & |  && | & & |  && | &\\
T(\dhh^{},\dkk^{}) & \longleftarrow & \cdot & \longleftarrow & 
\cdot & \longleftarrow & \cdot & \longleftarrow & \cdot  & ...
\end{array}
\end{displaymath}
for in virtue of formulas \mref{dTLp}, \mref{indag} and \mref{dGou} we have
$(\dhh^{\uparrow \dagger},\dkk^{\uparrow \dagger})=(\dkk^{\uparrow}_{(1)},\dhh^{\uparrow}_{(2)})=
(\dhh_{(2)},\dkk_{(2)})$  and analogously 
$(\dhh^{\dagger},\dkk^{\dagger})=
(\dhh^{\uparrow}_{(1)},\dkk^{\uparrow}_{(1)})$
i.e.  
\begin{displaymath}
\begin{array}{l}
T(\dhh^{\uparrow \dagger},\dkk^{\uparrow \dagger})=T(\dhh,\dkk), \qquad
T(\dhh^{\dagger},\dkk^{\dagger})=T(\dhh^{\uparrow},\dkk^{\uparrow}).
\end{array}
\end{displaymath}

\section{Conclusions}
We have shown  that the Moutard equation
splits in the dicrete case into discrete Moutard
equation and adjoint  discrete Moutard
equation.  This structure should be taken into considerations, when a
discretization of systems of a physical \cite{SchiefI,NDS,DNS}
and of geometrical  \cite{Nie,DoliwaK} importance  is discussed.

\bigskip 

{\large \bf Acknowledgments}

I thank the organizers of the Euro-conference NEEDS 2001 in Cambridge
for invitation and support that enable me to attend the conference.

\end{document}